\documentclass[
reprint,
superscriptaddress,
 amsmath,amssymb,
 aps,
 prx,
longbibliography,
floatfix,
]{revtex4-2}
 
 \usepackage{float}
\makeatletter
\let\newfloat\newfloat@ltx
\makeatother

\usepackage[normalem]{ulem}
\usepackage{appendix}
\usepackage{blindtext}
\usepackage{multirow}    
\usepackage{graphicx}
\usepackage{mathtools}
\usepackage{dcolumn}
\usepackage{amsthm}
\usepackage{tikz}
\usetikzlibrary{quantikz2}
\usepackage{amsmath}
\usepackage{amsfonts}
\usepackage{amssymb}
\usepackage{ctable}
\usepackage{bm}
\usepackage{bbm}
\usepackage{color}
\usepackage{placeins}
\usepackage{afterpage}
\usepackage{comment}
\usepackage{wasysym}
\usepackage[ruled,linesnumbered]{algorithm2e}
\usepackage{algpseudocode}
\usepackage{hyperref}

\usepackage[utf8]{inputenc}
\usepackage{xcolor}
\usepackage[margin=0.76in]{geometry}
\usepackage{siunitx}
\usepackage{multirow}
\usepackage{amssymb}
\usepackage{wasysym}
\usepackage{bibunits}
\usepackage{tabularx}
\usepackage{booktabs}
\hypersetup{
    colorlinks=true,
    linkcolor=blue,
    urlcolor=blue,  
    citecolor=blue,
}

\usepackage[caption = false]{subfig}
\hypersetup{
    colorlinks=true,
    linkcolor=blue,
    urlcolor=blue,  
    citecolor=blue,
}

\makeatletter
\newcommand{\vast}{\bBigg@{4}}
\newcommand{\Vast}{\bBigg@{5}}
\makeatother

\makeatletter
\def\maketitle{
	\@author@finish
	\title@column\titleblock@produce
	\suppressfloats[t]}
\makeatother 

\theoremstyle{definition}

\newcommand{\decodewin}{W}
\usepackage{braket}

\begin{document}

\title{Error correction of transversal CNOT gates for scalable surface code computation}

\author{Kaavya Sahay}
\affiliation{Yale Quantum Institute, Yale University, New Haven, Connecticut 06511, USA}
\affiliation{Department of Applied Physics, Yale University, New Haven, CT 06520, USA}

\author{Yingjia Lin}
\affiliation{Duke Quantum Center, Duke University, Durham, NC 27701, USA}
\affiliation{Department of Physics, Duke University, Durham, NC 27708, USA}

\author{Shilin Huang}
\affiliation{Yale Quantum Institute, Yale University, New Haven, Connecticut 06511, USA}
\affiliation{Department of Applied Physics, Yale University, New Haven, CT 06520, USA}

\author{Kenneth R. Brown}
\affiliation{Duke Quantum Center, Duke University, Durham, NC 27701, USA}
\affiliation{Department of Physics, Duke University, Durham, NC 27708, USA}
\affiliation{Department of Electrical and Computer Engineering, Duke University, Durham, NC 27708, USA}
\affiliation{Department of Chemistry, Duke University, Durham, NC 27708, USA}

\author{Shruti Puri}
\affiliation{Yale Quantum Institute, Yale University, New Haven, Connecticut 06511, USA}
\affiliation{Department of Applied Physics, Yale University, New Haven, CT 06520, USA}

\date{\today}

\begin{abstract}
Recent experimental advances have made it possible to implement logical multi-qubit  transversal gates on surface codes in a multitude of platforms. A transversal controlled-NOT (tCNOT) gate on two surface codes introduces correlated errors across the code blocks and thus requires modified decoding compared to established methods of decoding surface code quantum memory (SCQM) or lattice surgery operations. In this work, we examine and benchmark the performance of three different decoding strategies for the tCNOT for scalable, fault-tolerant quantum computation. In particular, we present a low-complexity decoder based on minimum-weight perfect matching (MWPM) that achieves the same threshold as the SCQM MWPM decoder. We extend our analysis with a study of tailored decoding of a transversal teleportation circuit, along with a comparison between the performance of lattice surgery and transversal operations under Pauli and erasure noise models. Our investigation builds towards systematic estimation of the cost of implementing large-scale quantum algorithms based on transversal gates in the surface code.

\end{abstract}
\date{\today}

\maketitle

\section{Introduction} \label{sec:intro}

Quantum error correction (QEC) protects encoded logical quantum information from decoherence on the underlying physical qubits~\cite{shor_fault-tolerant_1996, calderbank_good_1996}. 
Recent experimental progress has led to landmark demonstrations of fault-tolerant (FT)  state preparation~\cite{egan_fault-tolerant_2021, acharya_suppressing_2023, sivak_real-time_2023, acharya2024quantumerrorcorrectionsurface}, repeated error correction~\cite{schindler_experimental_2011, chen_exponential_2021,da_silva_demonstration_2024, kim_transversal_2024}, and state teleportation~\cite{ryan-anderson_high-fidelity_2024} of encoded logical states. In a multitude of platforms, high-fidelity two-qubit operations are no longer strictly confined to two-dimensional nearest-neighbor interactions~\cite{huang_comparing_2023,bluvstein_quantum_2022}, opening up the possibility to implement 
high-rate quantum LDPC codes ~\cite{panteleev_asymptotically_2022, lin_good_2022,bravyi_high-threshold_2024} and concatenated codes ~\cite{yamasaki_time-efficient_2024, yoshida_concatenate_2024}. Beyond this opportunity, non-trivial connectivity can be employed for
 logical operations in the widely studied surface code (SC), a leading candidate for practical quantum error correction~\cite{trout_simulating_2018, bluvstein_logical_2024}.

In fixed-qubit architectures, the prominence of the surface code can be attributed to its 2D planar layout, nearest-neighbour connectivity, low-depth stabilization circuits, and high error tolerance threshold~\cite{fowler_surface_2012,tomita_low-distance_2014,fowler_proof_2012}. Efficient graph-based decoders, such as minimum weight-perfect matching (MWPM), perform well at correcting common circuit-level errors~\cite{wu_fusion_2023,higgott_sparse_2023}. Further, logical gates on surface codes are well understood and easy to implement 
 with 2D nearest-neighbor connectivity via braiding~\cite{raussendorf_fault-tolerant_2007,fowler_surface_2012,brown_poking_2017} or lattice surgery~\cite{horsman_surface_2012,fowler_low_2018}. 
 
\begin{figure}
    \centering
    \includegraphics[width=\columnwidth]{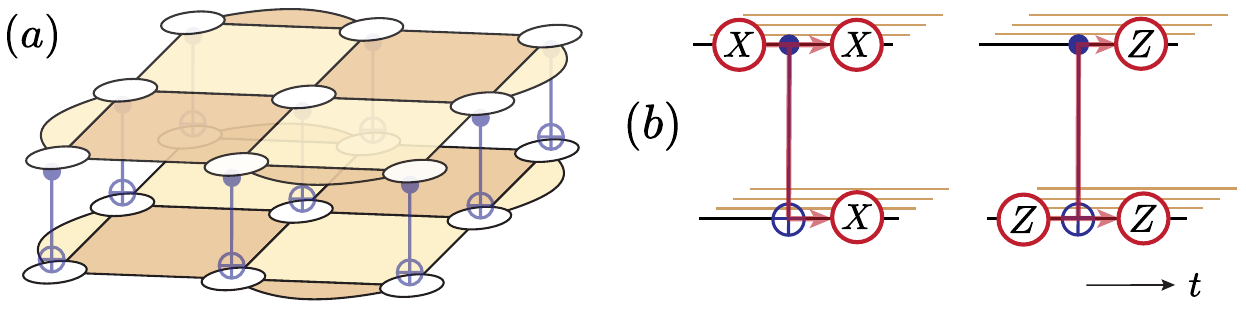}
    \caption{(a) A logical transversal CNOT operation between two rotated surface codes is performed by applying physical CNOT gates between each corresponding pair of data qubits of the SC states. (b) The transversal CNOT creates correlated errors between surface codes. Each SC is shown as a qubit set on which an error on a physical qubit can propagate to the other SC through the tCNOT.}
    \label{fig:tcnotErrs}
\end{figure}

 With non-local connectivity, it is possible to implement transversal logical gates, such as the logical CNOT (tCNOT), between any pair of surface codes. As shown in Fig.~\ref{fig:tcnotErrs}, this requires applying physical CNOT gates between every corresponding data qubit of the control and target SC states~\cite{horsman_surface_2012}. A tCNOT creates correlated errors. For example, a bit-flip error on a data qubit in the control may be copied over to the corresponding data qubit in the target. 
 One method to account for these error correlations 
 is by appropriately adding syndrome history from rounds following the CNOT gate from one SC to another~\cite{dennis_topological_2002,beverland_cost_2021}. Decoding using the resultant syndromes is suboptimal since the combined syndromes are twice as noisy as their individual components. 
 An alternative decoding strategy is to directly use all the measured syndromes of the two SCs without addition. In this case, decoding based on graph algorithms cannot be used and previous works thus resort to relatively slower hypergraph decoding~\cite{cain_correlated_2024,zhou_algorithmic_2024}.

In this work, we benchmark the performance of the transversal CNOT gate for scalable quantum computing using three decoding methods. In addition to the approaches described above, we study a thus-far uncharacterized strategy that we refer to as {\it ordered decoding}. In this approach, we first decode those errors in one SC state that may be copied over to the other SC state~\footnote{The basic concept behind ordered decoding was mentioned in a single sentence in~\cite{beverland_cost_2021}. However, there was no concrete implementation, analytical, or numerical results provided.}. We then correct for the identified errors on the second SC state before independently decoding the residual errors.
Any decoder that independently locates bit- and phase-flip errors can be employed in this overall strategy; here, we use MWPM. We find that ordered decoding using MWPM is highly effective at correcting noise correlations introduced by the transversal CNOT on surface codes, outperforming 
previous tCNOT decoders
in terms of thresholds. Our analysis extends beyond previously-studied constant depth circuits~\cite{viszlai_architecture_2023, ryan-anderson_high-fidelity_2024, cain_correlated_2024, zhou_algorithmic_2024, wan_iterative_2024}.

We additionally study decoding for transversal teleportation circuits in which one of the code blocks is measured soon after a tCNOT gate. Such teleportation circuits comprise a high fraction of two-qubit gate usage in quantum algorithms. We find that such teleportation operations can be inherently decoded using graph-based methods. We also provide a 
comparison for logical operations performed transversally versus via joint-parity measurements, i.e., lattice surgery, thus far the more widely studied method for surface code logical gate operations. This analysis is presented for Pauli-noise and a mix of erasure and Pauli noise. The latter noise model is motivated by 
recent studies showing that qubits with dominant erasure noise exhibit high thresholds and improved error-correction properties
~\cite{grassl_codes_1997, stace_thresholds_2009, wu_erasure_2022,kang_quantum_2023, teoh_dual-rail_2023, sahay_high-threshold_2023}.

Our work is structured as follows. In Section~\ref{sec:foundations}, we provide a brief introduction to the surface code when used as a quantum memory (SCQM), followed by an analysis of methods to decode and correct errors in this system. We then move onto logical computation using surface codes. We set out definitions for tCNOT circuits and argue for individually fault-tolerant tCNOT gadgets in logical algorithms in Section ~\ref{sec:scalable}. In Section~\ref{sec:tCNOTdecoders}  we discuss different tCNOT decoding strategies. In  Section~\ref{sec:tele}, we provide an analysis of gate teleportation, with a focus on decoding optimizations and a brief comparison with lattice surgery. Finally, in Section~\ref{sec:biaserase}, we discuss the performance of transversal and lattice-surgery based logical gates for erasure-based noise models. We conclude in Section~\ref{sec:conclusion}.

\section{The surface code as an error-corrected memory} \label{sec:foundations}

\begin{figure}
    \centering
    \includegraphics[width=\columnwidth]{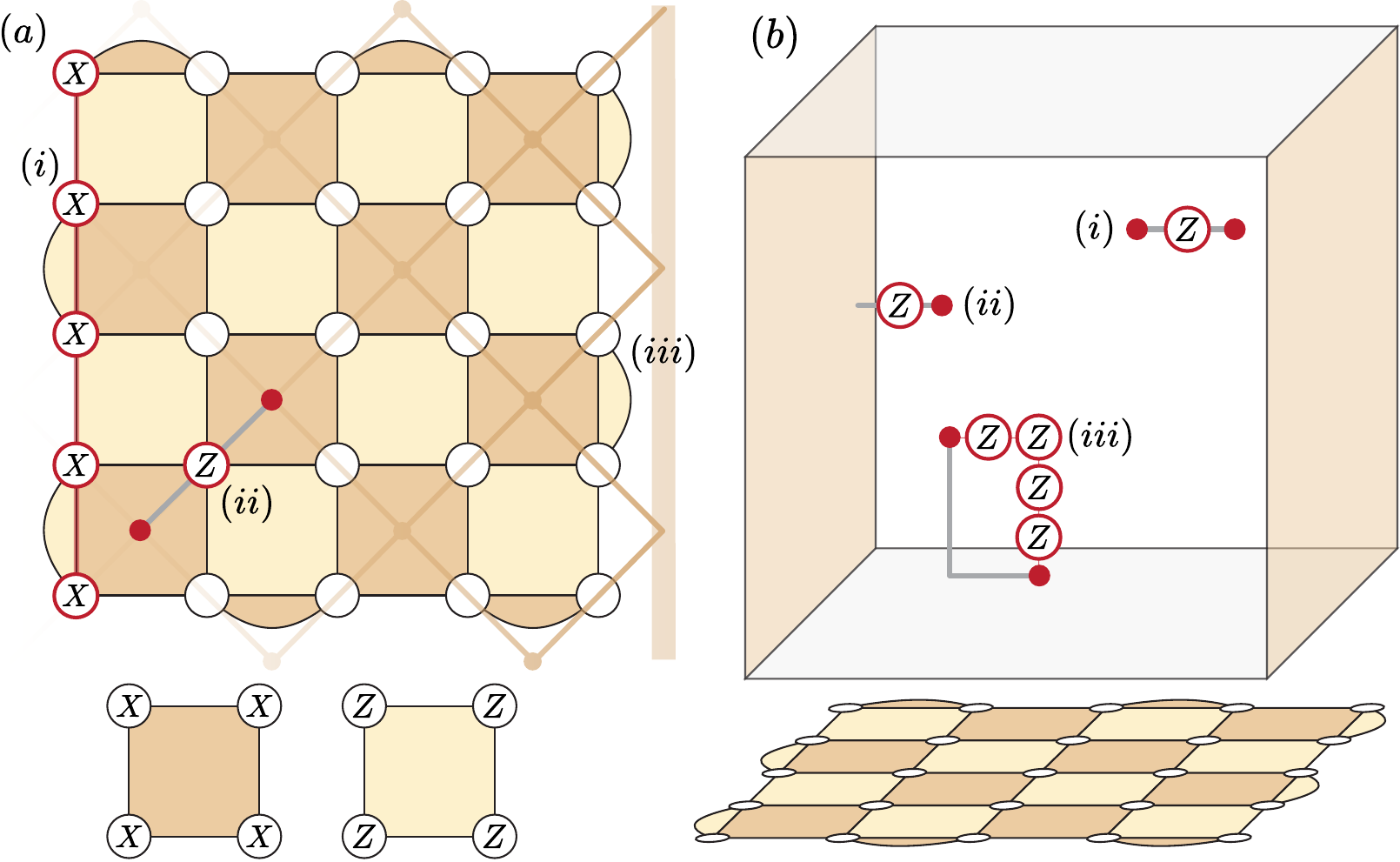}
    \caption{(a) A $d=5$ rotated surface code. (i) An $X$-logical operator, (ii) a single-qubit $Z$ error, with the corresponding anticommuting stabilizer measurements highlighted, and (iii) the X-decoding graph $G_X$ used to correct for $Z$ errors for one stabilizer measurement round. (b) A representation of $G_X$ generated by using $d$ rounds of stabilizer measurements on the underlying surface code. (i) Errors in the bulk create two defects to be matched together, (ii) An  error at the boundary creates a single defect, and (iii) a string of data qubit and measurement errors (red) and its corresponding matching-obtained correction (grey); the correction restores the original logical state up to code stabilizers.}
    \label{fig:1sc}
\end{figure}

The rotated surface code \cite{bravyi_quantum_1998,  bombin_optimal_2007} is a  stabilizer error correcting code \cite{gottesman_stabilizer_1997} that uses $d^2$ physical qubits arranged on the vertices of a $d\times d$ square lattice to encode one logical qubit. The length of the smallest logical operator, or equivalently the minimum number of Pauli errors to cause an undetectable change in the logical state, is $d$. Here, we take $d$ to be odd.
The stabilizer group $\mathcal{S}$ of this code is generated by  $X$ and $Z$ type {\textit{checks}} $S_X$ ($S_Z$) on alternating faces of this lattice, as illustrated in Fig.~\ref{fig:1sc}(a). Each $X$ ($Z$)-check is a product of Pauli $X$  ($Z$) operators on the qubits around the face.   The $X$- and $Z$- type logical operators, $\overline{X}$ and $\overline{Z}$, consist of Pauli $X$ and $Z$ operators on qubits lying on strings connecting  the boundaries of the lattice such that 
{ $\{\overline{X}, \overline{Z}\} = 0$, $[\overline{X}, S] = 0$ and $[\overline{Z}, S] = 0$ for all checks $S \in \mathcal{S}$}.  Figure~\ref{fig:1sc}(a) shows a $d=5$ surface code with a logical operator $\overline{X}$ (marked (i)).

{In practice, each $X$ ($Z$) check of the rotated surface code is measured using a depth-4 circuit of CNOT (CZ) gates to entangle the relevant physical data qubits with an additional ancilla qubit that is then measured out~\cite{tomita_low-distance_2014}, giving rise to a set of measurement outcomes referred to as the \textit{syndrome}. Errors can occur within this circuit at any point. Detectable errors anticommute with a subset of checks and flip the corresponding measurement outcomes, creating \textit{defects}. For example, Fig.~\ref{fig:1sc}(a)(ii), shows a $Z$ error on a data qubit that creates two adjacent defects. In the presence of faulty measurements, each stabilizer is generally measured  $O(d)$ times~\cite{dennis_topological_2002}}. This error correction protocol implements the identity channel on the encoded qubit - as a result, an isolated surface code acts as a quantum memory.

\subsection{Efficiently decoding errors on the SCQM}
\label{ssec:scqmDec}

{Given an error syndrome $\sigma$, optimal decoding involves finding a correction that maximizes the probability of restoration to the original code state or finding the most probable logical error. For general systems, this can be a computationally hard problem~\cite{iyer_hardness_2015}. 
In this section,
we give an overview of simpler polynomial-time decoders used for surface code error correction. 

We begin by defining a decoding hypergraph $G=(V,E)$ on a surface code with errors $\mathcal{E}$. Each vertex $v^t_S \in V$ corresponds to a {\it detector}, where a detector refers to the parity between the measurement outcomes of a check at time $t-1$ and $t$. In other words, $v^t_S = S^{t-1} \oplus S^{t}$ for $S\in \mathcal{S}$ \footnote{We consider $S^0$ and $S^{d+1}$ to be initial and final perfect check measurements respectively}.
The \textit{defects} $\mathcal{D}$ generated by $\mathcal{E}$ are detectors with odd parity under $\mathcal{E}$. Every hyperedge $e\in E$ is a set of detectors, and 
is assigned a weight proportional to the logarithm of the total probability of an independent error that causes that set of detectors to take the value 1.  Given $\{G, \mathcal{D}\}$, a graph-based decoder's task is to find the most probable physical error that created $\mathcal{D}$. 

For general hypergraphs where hyperedges correspond to more than two detectors, this is a computationally hard problem~\cite{berlekamp_inherent_1978}. By making the simplification of finding a locally - as opposed to globally- optimal solution, it is possible to efficiently find a correction operator $\mathcal{C}$ whose syndrome matches $\mathcal{D}$; one such strategy is referred to as the hypergraph union-find (HUF) decoder \cite{delfosse_toward_2021} .

Another simplification is to reduce $G$ to a graph where  $|e|\leq 2, \forall e$. For the surface code, this is possible for all single-qubit $X$ ($Z$) errors since these create independent pairs of defects (or a single defect at the boundaries) on the $Z$($X$) type stabilizer set, seen in Fig.~\ref{fig:1sc}(a)(ii). As a result, the decoding hypergraph $G$ can  be split into two disjoint graphs, $\{G_X, \mathcal{D}_X \}$ and $\{G_Z, \mathcal{D}_Z \}$, that satisfy $|e|\leq 2$ $\forall e \in E$. Fig.~\ref{fig:1sc}(a)(iii) shows an example of $G_X$ for one measurement round, and the box in Fig.~\ref{fig:1sc}(b) represents an example of $G_X$ for $d$ measurement rounds that we refer to as the spacetime decoding volume.  $Y$ noise is decomposed into $G_X, G_Z$ as uncorrelated $X$ and $Z$ errors.
{A decoder can now identify the most probable physical error in polynomial time by mapping $\sigma$ to minimum-weight matching problems on $G_X, G_Z$. 

For a given correction $\mathcal{C}$ found by a decoder, a corresponding update is applied to the surface code, ideally restoring it to the original state (as in Fig.~\ref{fig:1sc}(b)(iii)), but potentially causing a logical error if the correction proposed is logically inequivalent to the original error. For an MWPM-based decoder applied to a circuit-level noise model with two-qubit gate errors, the threshold error rate is $p_t \approx 1\% ~$\cite{wang_surface_2011, gidney_stim_2021}. We find the corresponding HUF threshold to be $0.89\%$.

\section{A scalable transversal CNOT} \label{sec:scalable}

In a transversal multi-qubit logical operation, a physical qubit of one logical block interacts with at most one physical qubit of another logical block. This approach naturally preserves the effective code distance. Here, we focus on a logical transversal CNOT (tCNOT) between   
a control surface code block $C$ and a target block $T$, implemented with physical CNOT gates applied between
qubit $q_C$ in $C$ and $q_T$ in $T$, for every physical qubit  $q$.

A tCNOT can introduce correlated errors between $C$ and $T$ via two mechanisms. First, two-qubit errors can occur after each physical CNOT gate. Furthermore, errors prior to the tCNOT can propagate from one code block to another. Specifically, $Z (X)$ errors on qubit $q_T$ ($q_C$) existing prior to the tCNOT are copied onto $q_C$ ($q_T$).
Importantly, the number of these errors copied over by the tCNOT scales linearly with the number of operations prior to the tCNOT. For example, if $r$ stabilizer measurement rounds preceded the tCNOT then the number of copied errors on each physical qubit grows as $O(r)$. A successful decoder must be able to decode such correlated errors across the logical code blocks in the circuit. 

With tCNOTs, in principle it is possible to decode over an entire algorithm using $O(1)$ rounds of syndrome extraction per tCNOT with well-prepared logical ancilla states~\cite{steane_active_1997,knill_fault-tolerant_2004,knill_quantum_2005,zhou_algorithmic_2024}. However, here we consider decoding for tCNOT gates at scale. 
In particular, we focus on quantum algorithms with the number of gates scaling exponentially with $d$, that require a distance $d$ fault-tolerant gate set for successful implementation~\cite{gidney_how_2021,dalzell_quantum_2023}.

For such a circuit,  we define a tCNOT \textit{gadget} to comprise of $g = O(1)$ tCNOTs in a known configuration. We take each gadget to be followed by $r$ rounds of stabilizer measurements on involved code blocks. 
For simplicity, we consider $g=1$, i.e. a gadget consists of a single tCNOT, and we ignore single qubit gates. 
In the following, we discuss the requirements for $r$ in the context of scalable quantum computing. 
Note that in this deep circuit, information from final transversal logical measurements is not readily accessible for use in decoding.

\begin{figure}
    \centering
    \includegraphics[width=\columnwidth]{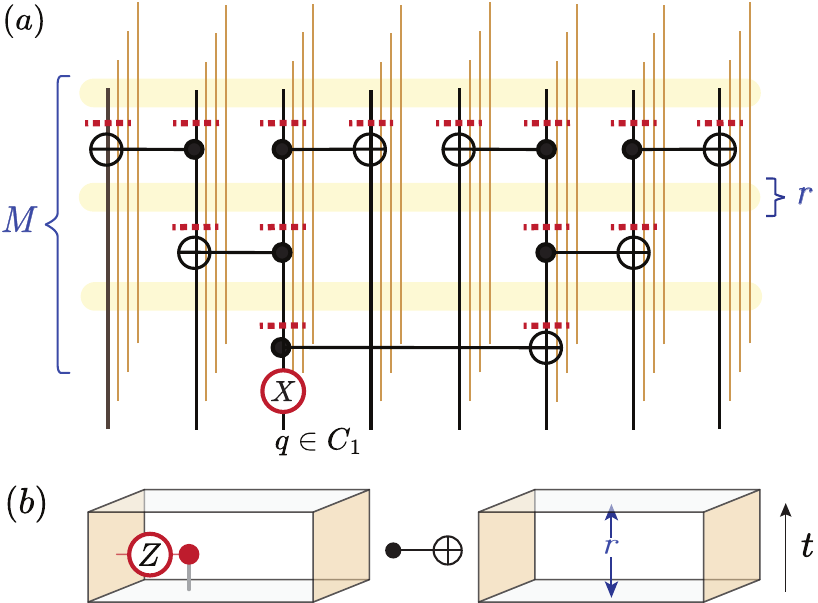}
    \caption{(a) A `binary-tree' tCNOT circuit of logical depth $M$. Each tCNOT is followed by $r$ rounds of stabilizer measurements (shaded yellow boxes). In this case a single error (for example an $X$ error on qubit $q$) can induce correlated errors that grow exponentially with circuit depth (marked by dashed red lines). (b) If $r << W$, each tCNOT cannot be decoded independently since weight $O(r)$ data errors can be misidentified as measurement errors, creating logical failures.}
    \label{fig:cnotCircs}
\end{figure}

We exemplify our arguments using a `binary-tree' circuit of logical gadget depth $M$ involving $2^M$ qubits.  Fig.~\ref{fig:cnotCircs}(a) shows an example circuit where $M=3$. In this circuit, a single bit-flip error $X_{q}$ on a qubit $q$ in the control SC of the first tCNOT, which we term $C_1$, can propagate to the corresponding qubits of all SCs. In a similar fashion, $Z$ errors on target SCs will flow to $C_1$. Correction of this first tCNOT should address both $X$ and $Z$ errors.

The surface code does not satisfy single-shot code properties for local check measurements~\cite{bombin_single-shot_2015, campbell_theory_2019, lin_single-shot_2023}, and decoding errors at a specific location requires roughly $\decodewin = O(d)$ subsequent rounds of stabilizer measurements~\cite{dennis_topological_2002,campbell_theory_2019, delfosse_beyond_2022}.  
Thus, if we set $r=\decodewin$, existing errors at each tCNOT  can naturally be fault-tolerantly corrected before the application of subsequent gates.
When $r \ll \decodewin$, however, one cannot decode each tCNOT  independently. 
To see this, consider Fig.~\ref{fig:cnotCircs}(b), 
with an open $Z$-error string $\mathcal{E}$ on $C$ right before the tCNOT connecting a spatial boundary to a single defect. 
Given a decoding window depth of $r$, a decoder can misinterpret $\mathcal{E}$ as a single length-$r$ measurement error string  if $|\mathcal{E}| > r$.
As a result, the decoder fails to correct $\mathcal{E}$.

When $r<\decodewin$, instead of attempting to decode individual tCNOTs, it may be possible to use a correlated decoding approach. Here, we use an expanded depth-$m$ ($m \le M$) window of the circuit extending over $2^m$ qubits. Syndromes in the entire window are used to decode errors at the beginning of the window~\cite{cain_correlated_2024,  zhou_algorithmic_2024}. 
This procedure is an extension of the overlapping- or sliding-window approach used for preserving a quantum state but applied to a quantum circuit block~\cite{skoric_parallel_2023,bombin_modular_2023, tan_scalable_2023,delfosse_spacetime_2023}. Note that in a general circuit, some qubits may be idle in the depth-$m$ window and these can be decoded separately as conventional quantum memories. 
In the rest of our discussion we will neglect these idle qubits.

\begin{table}[t]
    \centering \label{tab:rm}
\begin{tabular}{|c|c|c|c|c|c|c|}  
\hline
$r$ & $m$  & tCNOTs decoded    &  $C_1$ is & Decoding  \\
 &   & independently   &  FT &  volume \\
\hline
$O(1)$ & $O(1)$ & N  & N  & $O(1)$ \\

$O(1)$ & $O(d)$  & N   & Y & $O(e^d) $  \\

$O(d)$ & $O(1)$  & Y   & Y  & $O(d)$ \\
\hline
\end{tabular}
    
\caption{Constraints limiting tCNOT gadget separation $r$ and decoding window depth $m$ while decoding an exponentially growing tCNOT circuit. We assume $O(1) \ll d$.}
\end{table}

Let us examine correction of $X$ and $Z$ errors on  
$C_1$ in this window. Since $C_1$ is always the control of the tCNOTs, any error $X_q$ is copied from $C_1$ onto $2^m$ qubits, each of which may provide syndrome information about this propagated error. Conversely, $Z$ errors propagate \textit{onto} $C_1$ from $m$ logical gates. Decoding of  $Z$ errors on $C_1$ at the first tCNOT is dependent on decoding of the $2^m$ SCs that these errors can originate from. The total decoding volume for $C_1$, which determines the complexity of decoding in the correlated decoding approach, thus scales exponentially with $m$. 
If the decoding volume becomes too large, the backlog problem may be amplified~\cite{terhal_quantum_2015, higgott_sparse_2023}. Thus, we need to determine how this volume scales based on an $(r,m)$ choice that ensures that long-lived errors are prevented.

To this end, first consider $r,\; m\ll W$ for which, following previous arguments, we find that a decoder can fail to correct a $Z$-error string $\mathcal{E}$ of length $O(mr)$ on $C_1$ at the beginning of the window i.e., the first tCNOT. Thus, $r,m\ll W$ is not sufficient to prevent a long-lived $Z$ error. On the other hand, setting $mr=W$ is sufficient to prevent $Z$ errors at the beginning of the window from surviving for a long time. As expected, this condition cannot be satisfied if both $m, r = O(1)$. It can instead be satisfied by setting $r=O(1)$ and $m=O(W)$. In this case the decoding volume increases exponentially with $m$, leading to an undesirable slowdown in decoding. Clearly, the simplest way to achieve $mr=W$ without increasing the decoding volume exponentially is by choosing $r=O(W)$ and $m=O(1)$.  These conditions are summarized in Table I.

Hence we find that even in the expanded window approach $r = O(W)$ is desirable. Moreover, when $r = W$ and $m = 1$, each tCNOT gadget is effectively decoded independently. The advantage of these modular gadgets is that one can benchmark their individual failure rates and then estimate algorithmic performance for a wide variety of algorithms. For increasingly larger gadgets, the internal structure of the circuit determines how errors spread between different code blocks, complicating later circuit analysis. To predict algorithmic performance,  a defined set of gadgets would need to be benchmarked and the errors carefully composed.

Having addressed the effectiveness of $r=W$, we now discuss three decoding procedures that can be used to decode an individual tCNOT. Specifically we use the simplifying assumption of $\decodewin = d$ since measurement errors in our numerically simulated noise model are roughly as likely as data qubit errors.

\section{Decoders for a transversal CNOT}\label{sec:tCNOTdecoders}

\begin{figure*}[t]
            \centering
            \includegraphics
            [width=1\textwidth]{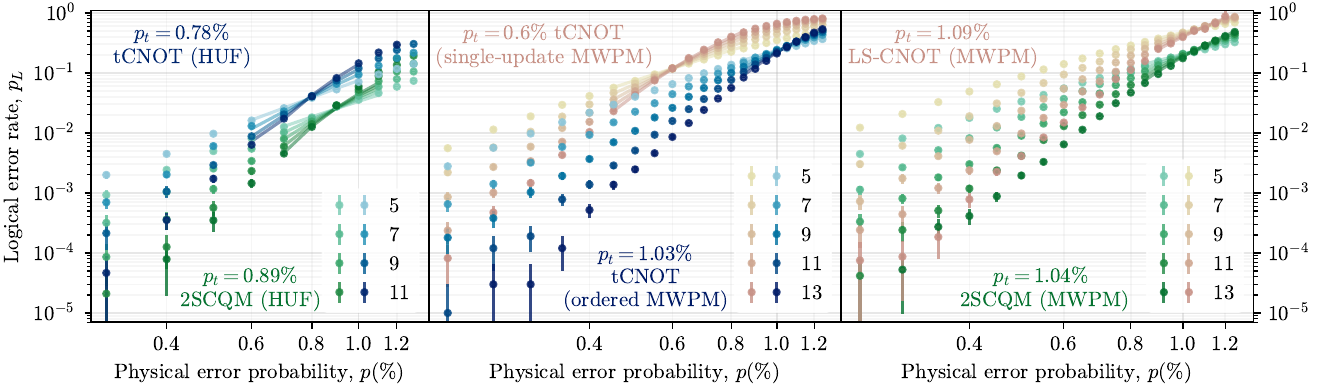}
            \caption{Thresholds and logical failure rates under circuit-level depolarising Pauli noise for SCQMs and logical CNOTs. (left) HUF decoder performance for a 2SCQM and tCNOT. (centre) single-update and ordered MWPM decoder performance for a tCNOT. (right) MWPM decoder performance for a 2SCQM and lattice-surgery (LS) CNOT.  Operations are plotted separately to prevent overcrowding. Thresholds are found by finite size scaling close to $p_t$ (translucent lines). }
            \label{fig:plotCNOTdecoding strats}
\end{figure*}

Here, we analyze decoding of a tCNOT. We consider a decoding volume comprising of the $d$ rounds of stabilizer measurements following the tCNOT along with the $d$ rounds of stabilizer measurements that followed the preceding gate. We focus on correcting $Z$-type errors using $X$-checks, using the notation $S_{CX}$ ($S_{TX}$) to refer to $X$-checks on the control (target); by symmetry, correction of $X$-type errors using $Z$ checks follows the same principles. 

For our decoding analysis, we introduce the unifying perspective of check \textit{frames} for SCs in the tCNOT. In what we term the \textit{dynamic frame}, each $X$ check used for correcting $Z$ errors evolves through the tCNOT 
as 
\begin{equation}
   \mathbf{S}_{CX}^i = S_{CX}^i  S_{TX}^i  \quad \forall i \in \{d+1,d+2,...,2d\}  \label{eq:cnotStabs}
\end{equation}.

This reflects the internal evolution of the checks caused by the tCNOT. $ \mathbf{S}^i_{CX} = S^i_{CX}, \forall i \leq d $ and $ \mathbf{S}_{TX} = S_{TX}$, i.e. other $X$-checks remain unchanged in the dynamic frame. In contrast, we define a \textit{static frame}, where all checks are left in their original values. 

For any given frame, we can define the $X$-decoding graph of the system $G$ as laid out in Sec.~\ref{sec:foundations}. For convenience in the chosen frame, we break up $G_X$ into \textit{subgraphs} $G_{CX}$ and $G_{TX}$, where  $G_{CX}$ ($G_{TX}$) contains the frame-defined nodes related to the checks of $C$ $(T)$. We term $G_{CX}$ a \textit{dependent subgraph} since its checks change due to the tCNOT in the dynamic frame. From a complementary perspective, the control can be regarded as dependent because $Z$ errors are copied over to $C$ by the tCNOT. Unlike $G_{CX}$, the nodes of $G  _{TX}$ are unaffected by the tCNOT; it is thus termed an \textit{independent subgraph}. 

Note that in  experiment, we always only measure checks in the static frame. Checks of the dynamic frame can be inferred by combining the outcomes of the static $X$-checks at the cost of making the inferred dynamic checks more unreliable than the measured static checks in the presence of measurement errors. }

\subsection{Single-update decoder} \label{ssec:suDec}

In the single-update decoding strategy, we operate in the dynamic frame. Post-tCNOT, since we update the $S_{CX}$ check measurement outcomes according to Eqn.~\ref{eq:cnotStabs},  the detectors  $
    v_{CX}^{{i}} =  \mathbf{S}_{CX}^{i-1} \oplus  \mathbf{S}_{CX}^{i}  $ correctly track checks through the tCNOT. 
We can now apply MWPM to decode the defects created in the new dynamic-frame detector set, similar to that of an SCQM. This is explicitly detailed in Alg.~\ref{alg:sudec}. Note that the use of the dynamic frame doubles the defect rate on $G_{CX}$ post-tCNOT. We provide examples of handling of relevant errors by the single-update decoder in Appendix~\ref{app:dynamicProp}.
 
To benchmark the performance of the single-update decoder, we compare the numerically calculated logical error rate for the tCNOT decoded in this fashion using MWPM, to that of two independent, disjoint SCQMs for $2d$ rounds of check measurements (hereon referred to as a 2SCQM experiment). The results are shown in  Fig.~\ref{fig:plotCNOTdecoding strats} centre and right. In this and all circuit-level simulations with Pauli noise, two-qubit gate errors are uniformly chosen at random from $\{I, X,Y,Z\} ^{\otimes 2} / \{I\otimes I\}$ at a rate $p$ while assuming no single-qubit gate, initialization or measurement errors~\cite{fowler_surface_2012} (see App \ref{app:sims}). As expected, due to the larger defect population on $G_{CX}$, this procedure results in a reduction in tCNOT threshold (${p_t} = 0.6\%$) compared to a 2SCQM (${p_t} = 1.04\%$). We note that the single-update decoding strategy was proposed in Ref.~\cite{dennis_topological_2002} and previously used in Ref.~\cite{beverland_cost_2021}.

\subsection{Combined hypergraph decoder}
We now move onto our second decoding strategy. The dynamic frame, while natively following the tCNOT induced check evolution, increases the number of errors on the dependent subgraph. It is thus natural to attempt to use the static frame where possible. As explained in Appendix~\ref{app:hyper}, it is possible to \textit{only} use the dynamic frame at the tCNOT round, with all other rounds in the static frame to define a set of detectors. 
We call this the \textit{hybrid frame}. In this frame, the tCNOT creates non-decomposable \textit{hyperedges} in the decoding graph, i.e. a single independent error mechanism creates three or more defects. This has been shown in Refs. ~\cite{cain_correlated_2024,zhou_algorithmic_2024, hetenyi_creating_2024}. A naive matching decoder cannot successfully decode hyperedges; we thus use a hypergraph decoder, specifically the hypergraph union find (HUF)  decoder \cite{delfosse_toward_2021,delfosse_almost-linear_2021, wu_wuyue16pkugmailcom_mwpf_nodate}.

We compare the performance of the tCNOT decoded with HUF to that of 2SCQM in Fig.~\ref{fig:plotCNOTdecoding strats} (left).  Interestingly, we see a slight reduction in threshold for the tCNOT (${p_t} = 0.78\%$) compared to the 2SCQM experiment (${p_t} = 0.89\%$). This may be attributed to the locally -- as opposed to globally --  optimal corrections found by the HUF  algorithm in combination with the increased complexity of hyperedges in the decoding graph for the tCNOT versus the 2SCQM.}

Hypergraph decoding presents a potential path to decode and correct logical operations that induce hyperedges. However, as seen above, time-efficient hypergraph strategies underperform with increased hypergraph complexity and have higher runtime overheads than their graph-based counterparts. Ideally, we would like an efficient decoding algorithm for the tCNOT that scales equal to or better than the equivalent decoder applied to a 2SCQM, while at the same time preserving the SCQM threshold. Our next decoding strategy, ordered decoding, achieves this goal.

\subsection{Ordered decoder} \label{ssec:orderedDec}

Let us now describe an ordered decoding strategy. This operates entirely in the static frame. We know that $Z$ errors occurring on the target pre-tCNOT -- and \textit{only} these $Z$ errors -- are propagated to the control. If we can, to the best of our ability, fully identify such errors via decoding the target first, we then know exactly what defects they will create on the control at round $d+1$. 

Ordered decoding relies on exactly this principle: we first decode $T$ using $G_{TX}$ in the static frame. From the resulting solution, the decoder identifies error clusters on $T$ that occur before the tCNOT. Detectors in $G_{CX}$ in the static frame that correspond to the propagation of these identified errors from target to  control at round $d+1$ are then flipped, keeping track of the updated logical status. This process changes the collection of apparent defects on the control. We finally simply decode and correct the control with the updated defect collection. A corresponding mechanism can also be applied to the $Z$-decoding graph with the control decoded before the target (see Alg.~\ref{alg:odec} for a full description).

 An ordered decoding strategy on a tCNOT results in the preservation of the SCQM threshold (see Fig.~\ref{fig:plotCNOTdecoding strats} (centre) with ${p_t}= 1.03\%$), with only a marginal bounded increase in logical error rates compared to an equivalent decoder on a 2SCQM (see App.~\ref{app:ambiguousErrs} for further analysis). Note that ordered decoding doubles the decoding time as decoding $C$ for $Z$ errors can only begin after the target is decoded (and vice-versa for $X$ errors). This constant factor increase does not amplify the backlog problem.

\section{Logical state teleportation\label{sec:tele}}

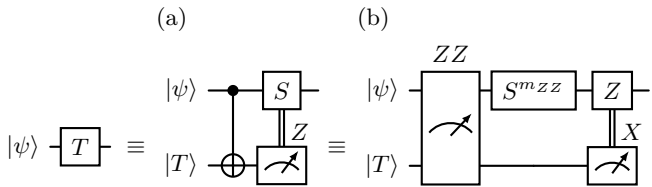
\begin{figure}[t]
    \centering
\begin{tabular}{lll}
 & (a) & (b) \\
\begin{quantikz}[column sep=0.14cm]
 \lstick{$\ket{\psi}$} & \gate{T}& 
\end{quantikz}  $\equiv$
&
\begin{quantikz}[column sep=0.16cm]
 \lstick{$\ket{\psi}$}   & \ctrl{1}  & \gate{S} \wire[d][1]{c} &  \\ 
 \lstick{$\ket{T}$}  & \targ{}    & \meter[label style={label={right:$Z$},inner sep=5pt,anchor=south east}]{} \\ 
\end{quantikz}  $\equiv$
&
\begin{quantikz}[column sep=0.16cm]
 \lstick{$\ket{\psi}$}  & \meter[2]{ZZ} & \gate{S^{m_{ZZ}}} &   \gate{Z} \wire[d][1]{c} &  \\ 
 \lstick{$\ket{T}$} &  &    & \meter[label style={label={right:$X$},inner sep=5pt,anchor=south east}]{} \\
\end{quantikz}
\end{tabular}
\caption{Methods to perform a $T$ gate using fault-tolerant magic state state teleportation. (a) A logical state is teleported onto the original qubit using a CNOT, followed by a Clifford update. (b) In a lattice-surgery setting, the equivalent protocol in (a) can be optimized to reduce the number of joint parity measurements~\cite{litinski_game_2019}. } \label{fig:Tcircs}
\end{figure}

One of the primary uses of two-qubit gates in quantum algorithms will be for logical state teleportation, particularly for non-Clifford gates, such as in Fig.~\ref{fig:Tcircs}. Here, we investigate these fault-tolerant  teleportation circuits, focusing on the teleportation step that may be implemented transversally or by joint-parity measurements. In a transversal teleportation circuit using a tCNOT, where a logical measurement is to be immediately implemented following the tCNOT, we find that, unlike the general unitary case,  subsequent stabilizer measurement rounds between the tCNOT and the logical measurement on the target  are not necessary to maintain code performance. We additionally compare a transversal teleportation strategy to that using joint measurements, an approach generally considered more suitable for planar architectures.

\subsection{Hypergraph reduction for transversal measurements}

 We first study the transversal teleportation scheme in Fig.~\ref{fig:Tcircs}(a), where a logical measurement of one of the surface codes immediately succeeds a tCNOT.  We consider $d$ rounds of stabilizer measurements on both the control and the target before the tCNOT. Post-tCNOT, since the logical $Z$ measurement on $T$ determines whether an $S$ gate is applied to the control, it is desirable that the logical $Z$ measurement is correctly read out. We find that the logical measurement terminates the decoding graph and reduces the hybrid frame hyperedges across the tCNOT to edges. Therefore, MWPM decoders are naturally applicable in this scenario.

 \begin{figure}
    \centering
    
    {\includegraphics[width=0.95\columnwidth]{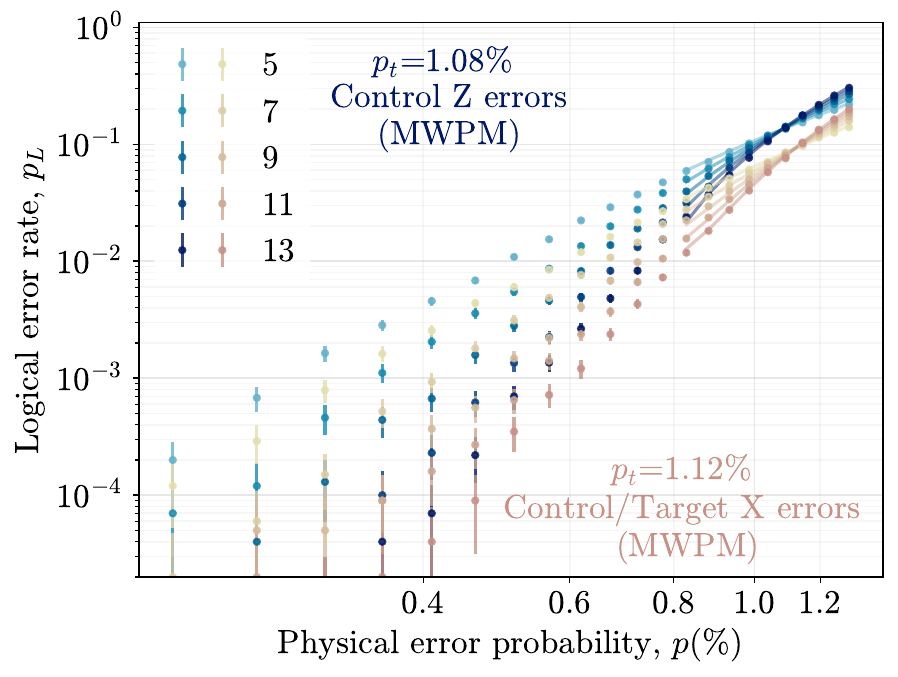}}\
    \caption{Thresholds and logical error rates for a gate teleportation circuit corrected using ordered decoding under circuit-level Pauli noise. We show the performance of target $X$ error (logical $Z$ measurement) and control $X$ error correction in one plot (lower points in brown) as they agree within error bars. Correction of $Z$ errors on the control is represented by the higher points in blue. Thresholds are found by finite size scaling close to $p_t$ (translucent lines). } 
    \label{fig:teleportation}
\end{figure}
 
We first focus on correction of the logical $Z$ measurement on the target. From the measurement results of all data qubits of $T$ in the $Z$ basis after the tCNOT, we can extract both the $Z$ logical measurement outcome and an extra $(d+1)$-th round of $Z$ stabilizer measurements from products of measurement results. 
 We use detectors of the hybrid frame (App.~\ref{app:hyper}).
 Crucially, unlike the tCNOT circuit we study in Sec~\ref{sec:tCNOTdecoders}, we do not include $d$ rounds of stabilizer measurement results on the control following the tCNOT during this correction. Therefore, there are no additional detectors  $v^i_{CZ}$ used for $i>d$ after the tCNOT. A measurement error on $S^d_{CZ}$ thus flips only two detectors $v^{d}_{CZ}$ and $v^{d+1}_{CZ}$ instead of three as described in Appendix~\ref{app:hyper}.
 Since each check is now involved in at most two detectors, this detector subset reduces hyperedges to edges. 
 For correction of the logical $Z$ measurement on $T$ with an MWPM decoder, we obtain $p_t=  1.12\%$, as shown in Fig.~\ref{fig:teleportation}. 

 We now move on to $X$ error correction  on $C$. During  correction of the logical $Z$ measurement on $T$, we  also obtain a correction for pre-tCNOT control $X$ errors. We first apply this correction to $C$ in an ordered decoding manner before continuing to measure $d$ rounds of stabilizers, which can subsequently be treated as an SCQM. Fig.~\ref{fig:teleportation} shows the logical error rate in this case, which overlaps with that of the target logical $Z$ measurement.
 
For 
$Z$ error correction on $C$ using $X$ checks,
we similarly find no hyperedges. Thus, graph-based decoding and its inherent logical error rates and thresholds are maintained when specific detectors are not present, as is in the case of transversal logical measurements immediately after a tCNOT.

\subsection{A comparison with lattice surgery} \label{sec:comparison}

In previous sections, we have laid out our scalable tCNOT strategy. In this section, we turn to a comparison with lattice surgery~\cite{horsman_surface_2012,fowler_low_2018, litinski_lattice_2018, litinski_game_2019,chamberland_universal_2022}. In lattice surgery, static logical surface code patches are set up with bridging regions of unentangled qubits between them. Stabilizer measurements on these bridging regions are turned on and off to connect the logical operators of individual surface codes and perform joint logical Pauli measurements. 
Reliable measurement of these joint Pauli operators requires $d$ rounds of stabilizer measurements. Arbitrary logical Clifford gates can be executed via combinations of these joint measurements and Pauli gates. 

The decoding of these operations has been well studied, being nearly identical to decoding a SCQM. As in the case of a SCQM, a conventional MWPM decoder works well, and thus we use this decoder for comparison.  We refer the reader to Appendix ~\ref{app:ls}-\ref{app:LSSA} for details. In practice, a lattice-surgery based quantum algorithm is compiled into the shortest sequence of joint measurements instead of directly using a gate set comprising of CNOT gates~\cite{litinski_game_2019, fowler_optimal_2013}, and a transversal implementation may use logical blocks with $m,g>1$. In this context, we leave the somewhat artificial comparison of an isolated fault-tolerant  tCNOT versus a lattice-surgery CNOT to  App.~\ref{app:lscnot}, showing therein that the tCNOT has both lower overheads and logical error rates. We display the numerical results in Fig.~\ref{fig:plotCNOTdecoding strats}. Here, we compare the cost of a state teleportation circuit implemented using both approaches that may be directly used in a logical quantum algorithm. 

In Fig.~\ref{fig:Tcircs}(b), we show the joint-parity version of the  teleportation circuit in Fig.~\ref{fig:Tcircs}(a). Since only a $ZZ$ logical measurement between the two surface codes is required, this operation takes $d$ rounds and requires no additional logical ancilla patches, which matches the overhead of an isolated fault-tolerant tCNOT within a logical algorithm. This suggests that in certain settings, the two strategies may use equivalent resources. We leave an extended overhead analysis to future work, focusing on fault-tolerant thresholds for both approaches in the next section.

\section{Logical gates for erasure qubits} \label{sec:biaserase}

Until now, we have analysed logical gates under i.i.d. Pauli noise. We now move onto a corresponding analysis for qubits where the dominant errors include a form of structured noise called \textit{erasures}~\cite{grassl_codes_1997, wu_erasure_2022}. Erasures consist of detectable errors that result in the affected qubit being projected into the maximally mixed state. We further investigate a more tailored model consisting of \textit{biased erasures}~\cite{sahay_high-threshold_2023}, where detectable errors only happen from one half of the computational subspace. We review the biased erasure noise model in App.~\ref{app:BE}.

\begin{figure}[t]
            \centering
            \includegraphics
            [width=0.96\columnwidth]{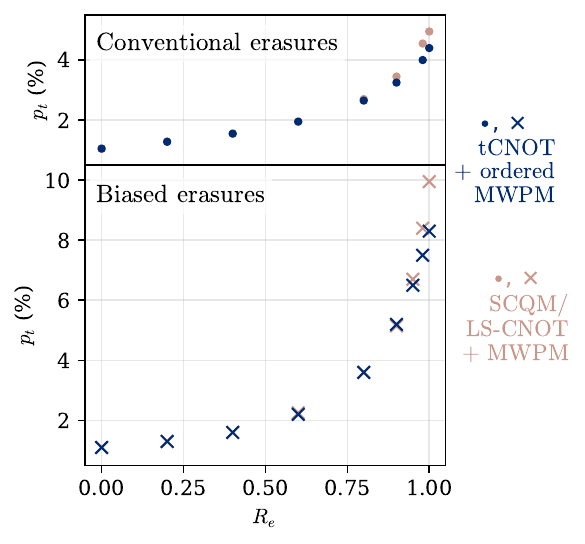}
            \caption{MWPM thresholds under circuit-level noise while varying the erasure fraction $R_e$ for a lattice-surgery CNOT (browns), and a tCNOT with  ordered decoding (blues). }
            \label{fig:threshVsRe}
\end{figure}

It is advantageous to engineer qubits whose dominant noise is erasures~\cite{wu_erasure_2022,kubica_erasure_2023}. In practice, not all noise can be converted to erasures; here, we assume the remainder to be depolarising Pauli noise within the computational subspace. This motivates us to define an erasure fraction $R_e$, i.e. given errors occurring on gates at a rate $p$ , $pR_e$ of them are converted to erasures, and the rest, $p(1-R_e)$ are Pauli errors. As different qubits operate at different erasure fractions, is instructive to see how the error correction properties of a code vary with changes in $R_e$. 

We analyse the change in MWPM-based thresholds with $R_e$ for logical operations in Fig.~\ref{fig:threshVsRe}. Fig.~\ref{fig:threshVsRe}(top) shows results for conventional erasures, and Fig.~\ref{fig:threshVsRe}(bottom) for biased erasures. We briefly summarize the results for biased erasures in the following: the threshold for a tCNOT corrected using ordered decoding increases from $1.03\%$ at $R_e=0$ to $8.3\%$ at $R_e \approx 1$. The thresholds for joint-measurements (denoted by LS-CNOTs in Fig.~\ref{fig:threshVsRe}) exactly coincide with a SCQM up to error bars, increasing from $1.04\%$ at $R_e=0$ to $10.3\%$ at $R_e \approx 1$. These values \textit{also} match with those of ordered decoding on a tCNOT except at extremely high erasure fractions $R_e>0.95$. Concretely, at $R_e=0.98$ we see that $p_t = 7.5\%$ for ordered decoding, while $p_t=8.3\%$ for regular decoding used during lattice surgery. A brief explanation for this phenomenon is given in Appendix~\ref{app:orderedErasures}. The fact that tCNOT thresholds for dominant erasure noise are strictly lower than the corresponding joint-measurements thresholds will ultimately affect the relative error rates of erasure-dominated transversal logical operations.

Similar behaviour is seen for conventional erasures, where $p_t = 4.4\%$ ($p_t = 5.0\%$) at $R_e \approx 1$ for ordered decoding on tCNOTs (decoding an SCQM). Note that conventional erasure thresholds at high $R_e$ are approximately half that of biased erasures. The significant increase in threshold at high $R_e$ for all logical operations considered, and the improved performance of biased erasures in comparison to other structured noise models demonstrates that the $ \text{Pauli} < \text{erasure} < \text{biased erasure}$ hierarchy of suppressed logical error rates and reduced hardware requirements demonstrated for SCQMs~\cite{sahay_high-threshold_2023} are retained for logical operations.

\section{Conclusion} \label{sec:conclusion}

In this work, we have performed an analysis of error correction of transversal CNOTs (tCNOTs) on surface codes in the context of scalable quantum computation. We highlight the utility of gadget fault-tolerance in large-scale quantum algorithms. In this context, we present a unified framework to describe various decoding strategies that may be used to correct tCNOT operations, focusing on an intuitive strategy, ordered decoding, that uses the deterministic propagation of Pauli errors through the logical gate. This strategy is compared with previous proposals to correct transversal logical operations - combined hypergraph decoding, and single-update decoding, showing that ordered decoding maintains the graph-based error correction advantages and thresholds of surface code memory experiments.

We extend our analysis in several ways: we study the special case of tCNOTs used for logical state teleportation, showing that the resultant hypergraph reduces to a graph in this instance, allowing regular surface code decoders to be used. We next perform a comparison of the transversal approach versus the joint-measurement based lattice surgery strategy, noting the possible reduced overhead of the former, with the caveat that the provided analysis is somewhat artificial given the differing optimal compilation schemes of the two strategies. Finally, we present an analysis of error correction of CNOTs under erasure-based noise models, showing a resultant increase in thresholds for dominant erasure noise for transversal and lattice-surgery based logical operations. 

We now briefly turn to a discussion focusing on hardware limitations. Most metrics we study indicate that where permitted by the architecture, a transversal strategy is more advantageous: tCNOTs corrected using ordered decoding have similar thresholds and lower logical error rates compared to lattice surgery-based approaches. There are, however, some subtle caveats. Transversal CNOT gates intrinsically involve nonlocal connectivity. Realistically, long-range interactions for a particular hardware may be slower and/or exhibit higher error rates than nearest-neighbour interactions. Specifically for the example of reconfigurable neutral atoms, the  corresponding error contribution arises from qubit movement across distances scaling with the surface code size~\cite{bluvstein_quantum_2022, evered_high-fidelity_2023}. As another example, for static neutral atom systems with long-range interactions realised via the Rydberg blockade, two-qubit gate fidelity decays quasi-linearly with the gate range \cite{pecorari_high-rate_2024,poole_architecture_2024}, setting a maximum radius over which using such non-local connectivity is practical.  
Thus, the relative performance of these gate strategies will vary significantly based on hardware constraints.

Our work thus builds towards an analysis of fault-tolerant transversal logical operations for use in quantum algorithms to be implemented in hardware. Further study of transversal non-Clifford gates~\cite{moussa_transversal_2016, brown_fault-tolerant_2020}, decoders for logical gadgets, and optimal compilation schemes~\cite{fowler_optimal_2013} that use a unified circuit-decoder perspective will indicate the performance of these strategies as a whole. 

\section{Acknowledgements}

We are grateful to Yue Wu, Shraddha Singh, Pei-Kai Tsai, Qile Su, and Aleksander Kubica for helpful discussions. In particular, we thank Jahan Claes for insightful contributions during early stages of this project.  We acknowledge the Yale Center for Research Computing for use of the Grace cluster. This work was supported by the National Science Foundation (QLCI grant OMA-2120757). Any opinions, findings, and conclusions or recommendations expressed in this publication are those of the authors and do not necessarily reflect the views of NSF. After the completion of this work, we became aware of a related  decoder implementation investigating low-depth circuits that uses the same underlying principle as ordered decoding~\cite{wan_iterative_2024}.

\bibliography{SCtCNOTGates}

\appendix

\setcounter{equation}{0}
\setcounter{figure}{0}
\setcounter{table}{0}
\renewcommand{\theequation}{A\arabic{equation}}
\renewcommand{\thefigure}{A\arabic{figure}}
\renewcommand{\thetable}{A\arabic{table}}

\section{Propagated defects in the dynamic frame} \label{app:dynamicProp}

Here, we expand on the functioning of the single-update decoder. Recall that it operates in the dynamic frame. In terms of the measured static checks, we have the dynamic frame detectors 
    \begin{subequations}
        \begin{align}
  v_{CX}^{{d}}  = S_{CX}^{d-1} \oplus  S_{CX}^{d} ,   \\
            v_{CX}^{{d+1}}  = S_{CX}^{d} \oplus  S_{CX}^{d+1} \oplus S_{TX}^{d+1}  , \text{ and} \\
            v_{CX}^{{d+2}}  = S_{CX}^{d+1} \oplus  S_{TX}^{d+1} \oplus S_{TX}^{d+2}  \oplus  S_{TX}^{d+2}         \end{align}
    \end{subequations}   

We illustrate the behaviour of the decoder using  an instance of a data qubit error $Z_{T,q}$ on a qubit $q$ in $T$ that creates defects on $T$ at round $k$:
\begin{enumerate}
    \item If $k\leq d$:  After round $d$, $Z_{T,q}$ is copied over to the control, creating $Z_{C,q}$. Since $ \left[ Z_{T,q}  Z_{C,q},  \mathbf{S}_{CX} =  S_{CX} S_{TX} \right] = 0$, no defect is  created in the dynamic frame after the tCNOT despite the actual error being present. This behaviour differs from the static frame.
    \item If $k>d$: $Z_{T,q}$ is not copied over to $C$, but creates new defects on the updated $C$ due to the updated checks of the dynamic frame: $ \left[ Z_{T,q}, \mathbf{S}_{CX} \right] \neq 0$. These errors are not truly on the control, and the resulting `false defects'  simply  mirror defect patterns on the target. Note that these false defects appear in addition to defects created by errors on the target itself, resulting in an effective doubling of the defect rate post-tCNOT on $T$.
\end{enumerate}

We observe that use of the dynamic frame prevents propagated errors from creating defects on $C$, while creating false defects from errors that did not propagate. After independently decoding and correcting both subgraphs in the dynamic frame, we use knowledge of all the corrections applied to $T$ to apply a final logical update to $C$ and restore the code states (Alg. \ref{alg:sudec}).

\section{Hyperedges in the tCNOT} \label{app:hyper}

Having discussed how the tCNOT copies over $Z$ errors from $T$ to $C$ in the main text, here we address how, in a certain frame, the tCNOT induces  non-decomposable hyperedges, i.e. leads to certain independent error mechanisms creating more than two defects in the decoding graph $G$, such that they cannot be expressed as products of regular edges.  Our discussion is framed in the language of a Stim circuit intended for PyMatching~\cite{gidney_stim_2021, higgott_sparse_2023} that uses a decoding graph with components in both the dynamic and static frames, i.e. the graph exists in a \textit{hybrid} frame. This versatility is facilitated by the ability of users to define detectors in Stim. We use the notation of Sec.~\ref{sec:tCNOTdecoders}, focusing on $Z$ errors and $X$-checks. Our system has the following three types of user-defined detectors:

\begin{enumerate}
    \item Until the tCNOT, the detectors are  $
    v_{CX}^{{i}} = {S}_{CX}^{i-1} \oplus {S}_{CX}^{i}$ and $
    v_{TX}^{{i}} = {S}_{TX}^{i-1} \oplus {S}_{TX}^{i} $ 
for $i \in \{1,2,...,d\}$. These detectors apply to both the static and dynamic frames.
\item Since the tCNOT copies over errors from $T$ to $C$, it is natural to assume that if no error occurs between measurement rounds $d$  and  $d+1$, then $ S_{CX}^{d+1} =  S_{CX}^{d} \oplus {S}_{TX}^{d}$. These three measurement outcomes thus can be used to define a detector $v_{CX}^{d+1} = S_{CX}^{d+1} \oplus  S_{CX}^{d} \oplus {S}_{TX}^{d} $. Detectors on $T$ remain unchanged: $v_{TX}^{d+1} = S_{TX}^{d+1} \oplus  S_{TX}^{d} $ . These detectors exist in a dynamic frame that is defined in a slightly different basis to Sec.~\ref{sec:tCNOTdecoders}.

\item After the tCNOT, in each individual surface code, stabilizer measurements should remain unchanged under error-free execution. This allows us to propose: $
    v_{CX}^{{i}} = {S}_{CX}^{i-1} \oplus {S}_{CX}^{i}$ and $
    v_{TX}^{{i}} = {S}_{TX}^{i-1} \oplus {S}_{TX}^{i} $ 
for $i \in \{d+2,d+3,..,2d\}$. Note that for $C$, these detectors are different from those defined by check evolution in the dynamic frame, and are instead in the static frame.
\end{enumerate}

With this set of detectors in the hybrid frame, it is possible to extract independent error mechanisms that create three defects. For example, consider a measurement error on $T$ at round $d$. Assuming no other errors, $S_{TX}^d = 1$ and all other outcomes are $0$, meaning that the set of detectors  $\{v_{TX}^{{d}} , v_{CX}^{{d+1}}, v_{TX}^{{d+1}} \} = 1$  produce defects.  This 3-component term is a hyperedge.

\section{Graph decoders for the tCNOT} \label{app:algs}

We provide a brief summary of the decoding algorithms based on independent handling of $X$ and $Z$ errors described in Sec.~\ref{sec:tCNOTdecoders}. By abuse of notation, we define $G_{ind} = \{ G_{TX}, G_{CZ}\} $ and $G_{dep} = \{ G_{CX}, G_{TZ}\} $. For every edge $e$ of weight $w$ in a graph $G$, we define $p_e = \frac{\exp ({-w})}{1 +\exp ({-w}) }$  

\begin{algorithm}
\caption{Single-update decoder}
\label{alg:sudec}
\tcc{Change dependent graphs to dynamic frame}
\ForEach{${S}^i_{dep} \in G_{dep} \mid i > d$}
    {
        ${S}^i_{dep} \gets {S}^i_{dep}\oplus {S}^i_{ind}$\
    } 
\ForEach{${v}_S^{i}\in G_{dep}$}
    {
        ${v}_S^{i} \gets {S}^{i-1}_{dep}\oplus {S}^{i}_{dep}$\
    }
\tcc{Updating dependent subgraph edge weights}
    
\ForEach{ $ e= (v_{S_k}^i, v_{S_l}^j)  \in G_{dep}$}
    {   
         \If{$ \exists v_{S_k}^i \in e \mid i > d $}{\hspace{1.5pt} $p_e \; \gets p_e +  p_{e{G_{ind}}}  - 2p_e p_{e_{G_{ind}}}  $}
    }

\textbf{Decode} $\{G_{ind}, G_{dep} \}$ independently\

\tcc{Update $G_{dep}$'s logical status}

$  L^{G_{dep}} \gets L^{G_{dep}} \oplus L^{G_{ind}} $
\end{algorithm}

 We note a fundamental conceptual similarity between ordered decoding and the two-pass or iterative MWPM decoding methods~\cite{fowler_optimal_2013, yuan_modified_2022, paler_pipelined_2023, iolius_performance_2023}. These approaches have been used to improve SCQM decoding accuracy by leveraging probabilistic intra-SC correlations. In contrast, for ordered decoding, we make use of deterministic inter-SC error correlations induced by the tCNOT. It may be possible to combine the two approaches, along with other subroutines that make use of soft information such as belief propagation~\cite{higgott_improved_2023}, to further improve the performance of ordered decoding.

\begin{algorithm}
\caption{Ordered decoder}
\label{alg:odec}
\textbf{Decode} $G_{ind}$ independently, obtaining the set of matched defect pairs $\mathcal{E}^{G_{ind}}$\

$L' = 0$

\tcc{Flip nodes in dependent subgraphs}
\ForEach{\normalfont{unordered pair }$e = (v_{S_k}^i, v_{S_l}^j) \in \mathcal{E}^{G_{ind}}$}
    {
         \If{$ \exists v_{S_k}^i \in {e} \mid i \leq d $}{\hspace{1.5pt} $ v^{d+1}_{S_k,dep}  \gets v^{d+1}_{S_k, dep} \oplus 1  \newline  v^{d+1}_{S_l, dep}  \gets v^{d+1}_{S_l,dep}  \oplus 1 $ \newline
         $L' \gets L' + L(e)$}
    }
\textbf{Decode} $G_{dep}$ independently\

\tcc{Update $G_{dep}$'s logical status}
$  L^{G_{dep}} \gets L^{G_{dep}} \oplus L'$
\end{algorithm}

\subsection{A phenomenological threshold comparison}

Hypergraph decoders on surface codes show an advantage over MWPM in the presence of $Y$ errors. This is because they can directly treat $Y$ errors using hyperedges. In contrast, an MWPM decoder must decompose $Y$ errors into independent $X$ and $Z$ errors and graphs. This decomposition results in the loss of initial $Y = XZ$ correlations. Here, we attempt to control for this possible $Y$ error advantage of HUF during tCNOT decoding. 

We perform a comparison between HUF and ordered MWPM decoding for a tCNOT using a tailored phenomenological  noise model. In this model, independent bit-flip errors on data qubits and measurement errors are applied with equal probability $p$ in each stabilizer measurement round. This removes $Y$ errors, allowing a direct comparison of the ability of the two decoders to correct correlated errors that arise solely from error propagation through tCNOT gates. 

\begin{figure}
{\includegraphics[width=0.98\columnwidth]{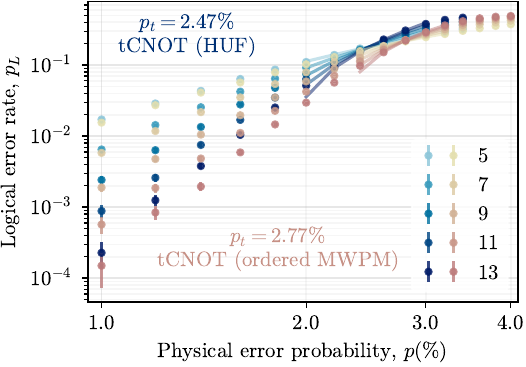}}\
    \caption{Thresholds and logical error rates of an HUF and ordered MWPM decoder for a tCNOT under biased phenomenological noise. Thresholds are found by finite size scaling close to $p_t$ (transculent lines).  } 
    \label{fig:phenomZ}
\end{figure}

We study the same tCNOT system as in Sec.~\ref{sec:tCNOTdecoders}, with the results shown in Fig.~\ref{fig:phenomZ}. Similar to the circuit-level model, we find a higher threshold using an ordered MWPM decoder ($p_t = 2.77\%$) compared to HUF ($p_t = 2.47\%$). Additionally, the logical error rate of ordered MWPM is slightly lower than that of HUF, reversing the general trend for circuit level noise that includes $Y$ errors seen in Fig.~\ref{fig:cnotCircs}. These findings suggest that (1) much of the logical error rate advantage of HUF indeed arises from its improved handling of $Y$ errors, and (2) an ordered MWPM decoder may outperform HUF when correcting correlated errors solely arising from tCNOT gates.

\section{Uncorrectable propagated errors for ordered decoding on the tCNOT} \label{app:ambiguousErrs}

\begin{figure}[t]
    \centering
\includegraphics[width=0.85\columnwidth]{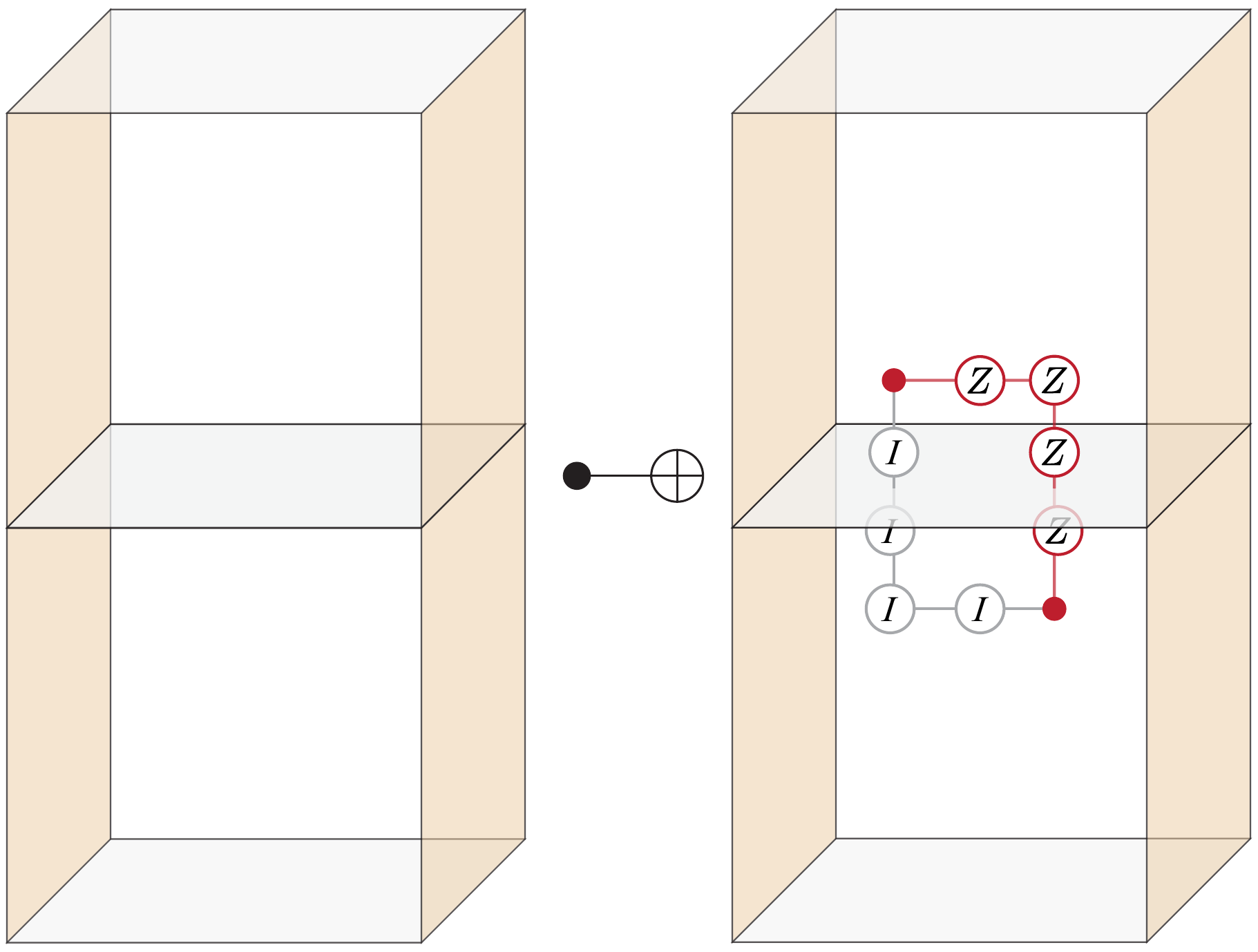}
    \caption{Errors can span the tCNOT measurement round, creating a defect below the tCNOT paired to one above it. Depending on where the data qubit errors actually occurred, different errors are copied over to $G_{dep}$ for the same defect pattern. If the error in red occurred, $C$ remains unaffected.}
    \label{fig:spacetimeCycle}
\end{figure}

 Here, we discuss the logical error rates observed for ordered MWPM decoding. For $Z$ errors, ordered decoding uses the simple concept of a target-first-control-next decoding strategy. The first round of decoding on $T$ pairs up defects on the $X$ stabilizer graph. These defect pairs correspond to errors that may or may not be copied over to $C$ by the tCNOT. Assuming errors are corrected up to stabilizers, these defect pairs arising from underlying physical errors, and their corresponding predicted  errors returned by a decoder create space-time cycles in the decoding graph (Fig.~\ref{fig:spacetimeCycle}). These cycles can be broadly classified into three distinct categories: 
 
\begin{enumerate}

     \item Spacetime cycles entirely before the tCNOT: The error corresponding to this defect pattern can be a combination of measurement and data errors. Only the  data-qubit part of this error, corresponding to its horizontal projection in spacetime, is transmitted to $C$. Even if the original error is corrected up to a time-like stabilizer, the dependent subgraph is correctly updated. 
     \item Spacetime cycles entirely after the tCNOT: The most probable error occurs entirely after the tCNOT and hence does not propagate onto $C$. 
    \item Spacetime cycles spanning the tCNOT round: This defect pattern arises from a combination of measurement and data errors. There are multiple minimal-weight ways of projecting such a path. For example, see Fig.~\ref{fig:spacetimeCycle}. Suppose the decoder identifies the grey path with the `data-qubit projection' entirely below the tCNOT, but the actual physical error was the red error with the `data-qubit projection' above the tCNOT. The former will transmit errors to the target, but the latter will not. Given the decoder's choice, the target will be updated incorrectly. Hereon, we refer to this class of errors as \textit{ambiguous} errors. 
    
\end{enumerate}

Ambiguous errors arising from spacetime cycles spanning the tCNOT round are effectively uncorrectable by the first stage of an ordered decoder - the decoder can only randomly choose the truly correct data-error projection. 
Subsequent error correction of the residual defects takes place on the dependent subgraph. As a result, ambiguous errors lead to an increase in the logical error rate of the dependent graphs. However, from Fig.~\ref{fig:plotCNOTdecoding strats}, we observe that they do not materially degrade decoder thresholds.

\subsection{Ordered decoding at high $R_e$} \label{app:orderedErasures}

Here, we discuss the apparent reduction in threshold at high erasure fractions for a tCNOT with ordered decoding compared to an SCQM experiment. In essence, this is because of  ambiguous errors that create cycles around the tCNOT round. For all other categories of correctable erasure errors, the decoder can use the erasure flag on qubits suffering errors to determine the exact path of errors propagating over onto $G_{dep}$.

Observe the example Fig.~\ref{fig:spacetimeCycle}. Unlike the previous section, we envision that erasure errors have occurred along both grey and red paths. However, only one (red) path truly has $Z$ errors due to the erasures, giving rise to two defects. Exactly like the case with Pauli errors, the decoder has no way to determine which of the two error patterns occurred. If it guesses the grey path occurred, an incorrect update is applied to the control. At high $p$ and $R_e$, the probabilities of such cycles increase. At sufficiently high $p R_e$ logical errors are dominated by contributions from these ambiguous cycles for which the decoding ability is Pauli-like instead of erasure-like. This in turn reduces $p_t$ for the tCNOT at high erasure fractions.

\section{Logical computation using lattice surgery} \label{app:ls}

An SCQM requires at least $d$ errors to create an undetectable logical operator. Ideally, this distance $d$ to errors should be preserved while using individual surface codes as units of logical computation. Logical gates executed via lattice surgery achieve this in a manner compatible with planar fixed qubit architectures~\cite{fowler_surface_2012,litinski_game_2019}.

\begin{figure}
    \centering
    \includegraphics[width=\columnwidth]{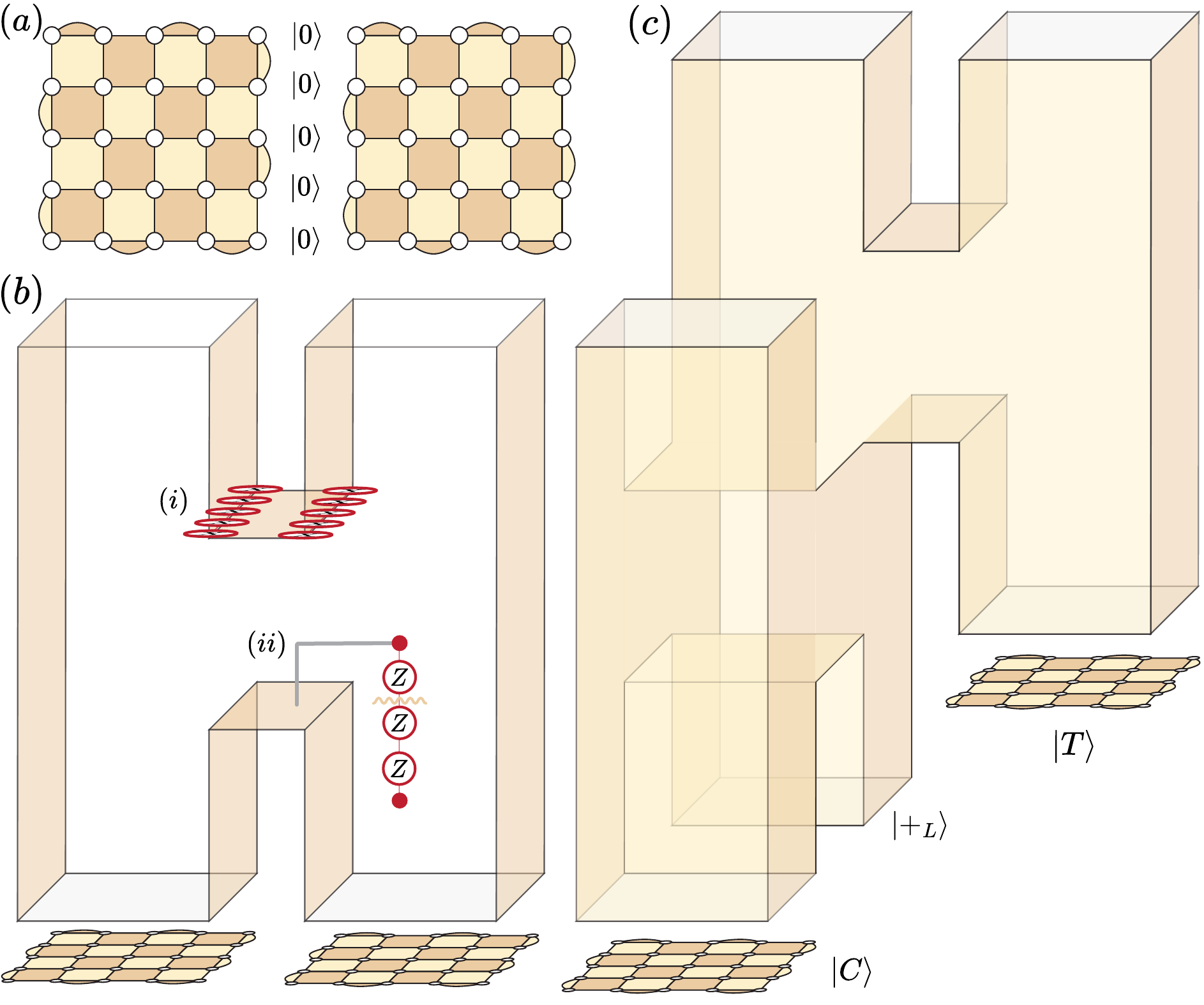}
    \caption{(a) Setup, and (b) spacetime volume of an $XX$ logical measurement conducted using lattice surgery, with (i) the product of stabilizers denoting the $XX$ measurement outcome, and (ii) an error that will be misidentified without the buffer measurement rounds. (c) Spacetime decoding volume for a logical CNOT done with lattice surgery.}
    \label{fig:2ls}
\end{figure}

We illustrate the process for a joint-Pauli $XX$ measurement done via lattice surgery on two surface codes in Fig.~\ref{fig:2ls}(a)-(b). The system is set up as follows: two initially disconnected surface codes have their individual stabilizers measured for $O(d)$ rounds. A  bridging region of width $b$ between the $X$-logical edges of these codes is initialized with unentangled qubits in $\ket{0}$ (Fig.~\ref{fig:2ls}(a)). To begin the joint-parity (JP) measurement, surface-code checks are measured over the entire region comprising of the two logical qubits and bridging region. This projects the two surface codes into a joint logical subspace. 

In the absence of errors, all check measurements in the logical region, and $Z$ stabilizer measurements in the bridging region return $+1$ outcomes. $X$ stabilizers in the bridging region are initially unfixed and yield random outcomes, but their combined product is the $XX$ logical measurement outcome (Fig.~\ref{fig:2ls}(b)(i)). To achieve  fault-tolerance to measurement errors that may lead to misidentification of the $XX$ measurement result, JP measurements are done for $O(d)$ rounds.  After these JP rounds, the bridging region qubits are measured in the $Z$ basis, and the surface codes are once again measured independently. The spacetime decoding volume of this process is shown in Fig.~\ref{fig:2ls}(b).

Extending this strategy, a logical CNOT gate between a control surface code $C$ and a target surface code $T$ can be performed using a similar JP-measurement based circuit. This  operation uses a logical $ZZ$ measurement on $C$ and a logical ancilla patch, followed by a logical $XX$ measurement on the ancilla and $T$.
In practice, the spacetime volume of this set of operations looks like Fig.~\ref{fig:2ls}(c) where $O(2d)$ round of gates along with one logical ancilla are used to perform the complete gate. Note that single-control $n$-target CNOT gates can be carried out simultaneously by using a large ancillary region that can support $X^{\otimes n}$ measurements \cite{fowler_low_2018,litinski_lattice_2018}. This CNOT operation has the same threshold as a SCQM, and the same logical error rate scaled up to known constant prefactors (discussed in App.~\ref{app:lsEC}). This is verified via numerical simulations in Fig.~\ref{fig:plotCNOTdecoding strats}.

\section{Error correction in lattice surgery} \label{app:lsEC}

We expand on decoding and correction of errors during the $XX$ joint parity measurement discussed in App.~\ref{app:ls}.   Close observation of the ‘legs’ of the H in Fig.~\ref{fig:2ls}(b) – representing pre-JP stabilizer measurement rounds on the individual surface codes, hereon referred to as \textit{buffer rounds} -  show that they are equivalent to the spacetime volume of individual SCQM experiments (up to time boundaries that connect them to the JP rounds). The bridging region in the joint parity rounds is a representation of the stability experiment \cite{gidney_stability_2022}.The stability experiment can be interpreted as  a space-time rotated version of the SCQM. 

As a combination of variously orientated SCQMs, we can observe that the complete decoding graph of the logical $XX$ measurement is essentially a trivially extended SCQM. As a result, in lattice surgery, we preserve the threshold of an equivalent SCQM. This has previously been demonstrated in Ref.~\cite{chamberland_universal_2022} using a matching decoder. Further, the total logical error rate of a lattice surgery operation can be expressed in terms of its logical spacetime surface area (LSSA) in comparison to an SCQM's LSSA, a concept we discuss in App.~\ref{app:LSSA}.

An additional note of interest is the choice to use buffer rounds at all; we connect this to the notion of $\decodewin$ in the main text. Technically, these legs of the ‘H’ shaped spacetime volume are not part of the JP measurement itself. Indeed, in the ideal scenario of perfect state preparation of the disjoint surface codes, or surface code readout immediately after the JP rounds, they are not required. However, measurement errors in the original logical patches can create defects that can be misidentified as faulty joint-parity measurement errors had measurement data from the buffer rounds not been used. Fig.~\ref{fig:2ls}(b)(ii) exemplifies this: if the decoding window does not extend below the first JP round, the lone defect above the window boundary is misidentified as arising from a JP $X$ stabilizer measurement error. Thus, we use $\decodewin$ buffer rounds of decoding data for fault-tolerance against these errors. Like the tCNOT, these buffer rounds can be part of  other non-commuting JP measurements -  inclusion of their measurement data is only necessary for decoding and they do not actually need to be perfectly disjoint rounds on individual surface codes (see Fig.~\ref{fig:2ls}(c)).

\section{Logical surface areas for SC operations}\label{app:LSSA}

For the operations we study, the failure rates of different logical gate strategies at low physical error rates can be approximately mapped to the spacetime surface areas ratios of their logical operators.

As an example, for a single SCQM, there are four logical operators of length $d$ (two $X$ logicals on opposite boundaries of the surface code and two $Z$ logicals). Each logical is measured for $d$ rounds, so the total logical spacetime surface area is $4 \times d \times d = 4 d^2.$ We count in units of $d^2$, and so this is simply 4 in our chosen unit system.

We now describe a 2SCQM, i.e. a memory experiment on two surface codes for $2d$ rounds, the logical spacetime  surface area (LSSA) of which is:.
\[I_{2SCQM} = \overbrace{2}^{\text{SCs}} \times \overbrace{2}^{t} \times \overbrace{4}^{\text{SA per SC}} = 16\]

Let us next examine an $XX$ logical measurement on two surface codes. The bridging region consists of a width $b$ strip that uses $bd$ qubits. Depending on the architecture, $b$ can be chosen to be constant or scale linearly with $d$. The two surface codes are measured independently for $d$ rounds, have projective measurements onto the $XX$ basis using the bridging region for $d$ rounds and are then split and measured again for $d$ rounds. The LSSA $XX_{2SC+3d}$ for this operation is given by:
\[\overbrace{2 \times (\overbrace{ 2\times 4}^{\text{split rounds}} + \overbrace{ 3}^{\text{merge rounds}})}^{\text{SCs for}} + \overbrace{4 \times b/d}^{\text{bridge}}  = 22 + 4b/d\]

\begin{figure}[t]
    \centering
\includegraphics{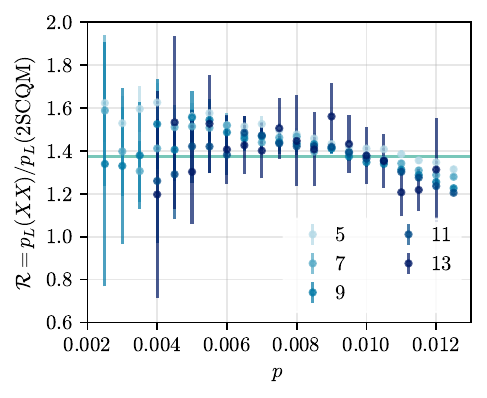}
    \caption{Ratio of logical failures rates for a $XX$ measurement using lattice surgery with $b=1$ to a 2SCQM. Each point uses $10^{5}$ samples, discarding points with failure rates $\leq 10^{-4}$. Error bars represent the 95\% confidence interval. A horizontal line is drawn at the lower bound of $1.375$.}
    \label{fig:XXvsMem}
\end{figure}

We can thus find the logical error rate ratio of a lattice surgery $XX$ measurement versus a memory experiment as:
\begin{equation}
    \frac{XX_{2SC+3d}}{I_{2SCQM}} = 1.375 + \frac{b}{4d}
\end{equation}
This is verified numerically in Fig.~\ref{fig:XXvsMem} which uses $b=1$. 

We can similarly calculate the LSSA for a logical CNOT between two surface codes using lattice surgery. As described in App.~\ref{app:lsEC}, this operation takes a total of $4d$ rounds. Its LSSA  ${CX}_{2SC+4d}$ is: 
\[ \overbrace{3 \times 2\times 4}^{\text{independent SCs}} + \overbrace{ 3 \times 4}^{\text{SCs while merging}} + \overbrace{4 \times 2 \times b/d}^{\text{bridge}}  = 36 + 8b/d\]

We can thus verify the logical error rate ratio of a lattice surgery CNOT versus a memory experiment as:
\begin{equation}
    \frac{CX_{2SC+4d}}{I_{2SCQM}} = 2.25 + \frac{b}{2d}
\end{equation}

\subsection{LSSAs for tCNOTs}

We now calculate the magnification of error rates induced by ordered decoding at low $p$. The overall failure rate of the tCNOT is the probability that the decoder applies a logically incorrect update to any of the four $\{ G_{CX}, G_{CZ} , G_{TX}, G_{TX}\} $ decoding subgraphs. For a 2SCQM, these four rates are equivalent. For a tCNOT with ordered decoding, we take the dependent subgraphs to have a 50\% increase in logical failure rates compared to a 2SCQM because they use their own LSSA combined with half the LSSA of their corresponding independent subgraph. The independent subgraphs remain unaffected. Thus, the logical rate amplifier is: $\frac{2 + 2\times {1.5} }{2+2} = 1.25$

\begin{figure}[t]
    \centering
\includegraphics{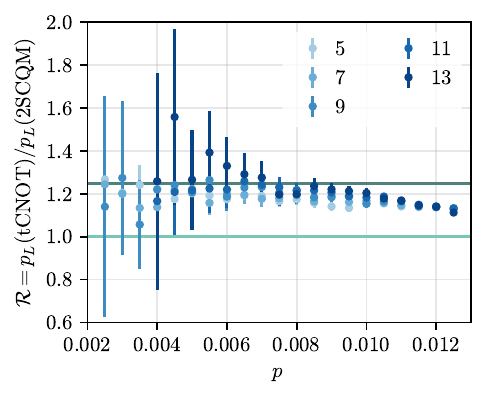}
    \caption{Ratio of logical failures rates for a tCNOT with ordered decoding to a 2SCQM. Each point uses $10^{5}$ samples, discarding points with failure rates $\leq 10^{-4}$. Error bars represent the 95\% confidence interval. Horizontal lines are drawn at $1$ and $1.25$.}
    \label{fig:orderedComp}
\end{figure}

In Fig.~\ref{fig:orderedComp}, we verify these calculations. We find the ratio $\mathcal{R}$ of the numerically obtained logical failure rates for a tCNOT corrected using ordered decoding to that of a 2SCQM for various system sizes. By virtue of incorrectly propagated logical updates,  ambiguous errors, and errors introduced within the faulty physical CNOT gates that implement the tCNOT, we observe that $\mathcal{R}$ is consistently higher than 1 and seems to saturate below threshold. By fitting to a saturation function, we find a saturation $\mathcal{R} \approx 1.25$ at low $p$, represented by the dark green horizontal line. This agrees with the preliminary analytical prediction above.

\subsection{Multi-target CNOTs}

We can now perform the same analysis for an $n$-target lattice-surgery CNOT, which can be executed in $2d$ rounds. Note that here the required number of extra bridging qubits is lower bounded at $\lceil n/2 \rceil bd$. We get the LSSA ${CX_\text{LS-n}}$:
\[ \overbrace{3 \times (n+1)\times 4}^{\text{independent SCs}} + \overbrace{ 3 \times (2 + (n+1))}^{\text{SCs while merging}} + \overbrace{4 \times 2 \times \lceil n/2 \rceil bd}^{\text{bridge}} \]
Comparing it to the identity operation,
\begin{equation}
    \frac{{CX_\text{LS-n}}}{I_{(n+1)SC+2d}} =
     \frac{21 + 15n + \lceil n/2 \rceil 8bd }{8(n+1)}
\end{equation}

Similarly, we evaluate the LSSA of an $n$ target tCNOT with ordered decoding. Note that this can either be executed natively in hardware, or using consecutive single-target tCNOTs, with the block then requiring a subsequent $\decodewin $ buffer rounds, without exacerbating the decoding complexity. We can then compare it to the identity operation,
\begin{equation}
    \frac{{CX_\text{t-n}}}{I_{(n+1)SC+2d}} =
     \frac{ (n+1) + n\times 1.5 + ( 1 +0.5n )  }{2(n+1)} 
\end{equation}

\section{An isolated LS-CNOT vs a tCNOT} \label{app:lscnot}

\begin{table*}[t]
    \centering
\begin{tabular}{|c|c|c|c|c|c|c|c|}
\hline
Method of conducting & No. of  & No. of  & Decoding time & Minimal patch & Amplification of logical \\
 logical CNOT & additional qubits & msmt rounds & complexity &  movement &  error rate over 2SCQM \\
\hline
Lattice surgery & $d^2 + 2bd$&  $2d$ & $O( \left(10d^3 + 2bd^2\right)^3)$ & Y & $2.25 + b/2d$ [App.~\ref{app:LSSA}] \\
tCNOT + ordered decoding & 0 & $d$  & $2 \times  O( \left(2d^3\right)^3)$  & N &  $\sim 1.25$ [App.~\ref{app:LSSA}]\\
\hline

\end{tabular}
\caption{A preliminary overhead comparison for an isolated fault-tolerant logical CNOT performed via lattice surgery vs. transversally with ordered decoding. We  use a MWPM decoding subroutine, resulting in both CNOT strategies having a threshold of $1\%$. We benchmark the number of additional qubits and measurement rounds needed for both strategies, along with the decoding time complexity and total logical error rate in comparison to a 2SCQM. }
    \label{tab:comparison}

\end{table*}

Here we discuss our results benchmarking the performance of a fault-tolerant tCNOT against a lattice-surgery-CNOT, in a setting where state preparation and measurement do not immediately occur before and after the logical operation. We highlight that this comparison is quite artificial, given that (1) in practice for transversal circuits, it may be more advantageous to use $m, g >1$ tCNOTs in a logical decoding block (see Sec.~\ref{sec:scalable}), and (2) in practice for joint-measurement based circuits, Clifford gates can be time-efficiently compiled~\cite{fowler_time-optimal_2013}. A more realistic analysis is left to future work.

We find that for a tCNOT, the total logical error rate is only marginally amplified compared to a 2SCQM experiment. However, it more than doubles for a LS-CNOT on account of that fact that it needs two logical Pauli measurements (see Appendix.~\ref{app:LSSA} for an analysis). Further, the lattice surgery approach requires an additional ancilla surface code patch, as well as bridging regions of uninitialized qubits. This results in a total additional qubit overhead of $d^2 + 2bd$, where $b$ is the width of the bridging region of qubits. $b$ is dependant on the architecture chosen. For movable architectures, this can be $O(1)$, but for fixed architectures this is generally the separation of the surface codes patches. These extra qubits are not needed for a tCNOT. Using a MWPM subroutine, the time overhead for ordered decoding of the tCNOT scales as $2 \times  O( \left(2d^3\right)^3)$. The factor of two arises since the independent and dependent graphs are decoded serially. This same quantity is $O( \left(10d^3 + 2bd^2\right)^3)$ for lattice surgery, implying that despite the inherent latency of ordered decoding, a correction $\mathcal{C}$ may be identified faster than an equivalent lattice surgery instance. The results of this comparison are summarized in Table ~\ref{tab:comparison}.

\section{Details of numerical simulations}
\label{app:sims}

The simulations based on MWPM decoders for the SCQM and  transversal CNOT operations were implemented in C++ using the Blossom algorithm~\cite{edmonds_paths_1965, kolmogorov_blossom_2009}, combined with Djikstra to account for adjusted edge weights arising from erasures. The code for these is available at \url{https://github.com/kaavyas99/CSS-tCNOT-decoders}. The corresponding HUF-based simulations were implemented in Stim~\cite{gidney_stim_2021}, and decoded using the MWPF implementation of  HUF~\cite{wu_wuyue16pkugmailcom_mwpf_nodate}. A second implementation of ordered decoding in Stim using PyMatching~\cite{gidney_stim_2021,higgott_sparse_2023} was additionally developed for further testing.

For the lattice surgery analysis, simulations and measurements of the logical error rate were conducted for a logical $XX$ measurement instance using the procedure detailed in Ref.~\cite{chamberland_universal_2022} in C++ using Blossom and Djikstra. The results reported for the CNOT were then calculated by scaling according to the operations' respective LSSAs.

In the noise model used for simulations, two-qubit gate errors are uniformly chosen at random from $\{I, X,Y,Z\} ^{\otimes 2} / \{I\otimes I\}$ at a rate $p$. This model also applies to the physical CNOT gates used in the logical tCNOT. State preparation, measurement, idling, and single-qubit gate errors are not included~\cite{fowler_surface_2012}. 

For data displayed in Figs.~\ref{fig:plotCNOTdecoding strats} and \ref{fig:teleportation}, individual  points are collected
using $10^5$ Monte Carlo samples; error bars are found by jackknife resampling and represent the 95\% confidence interval. Note that we report the total logical error rate per operation, i.e. over $2d$ measurement rounds for tCNOTs, as opposed to the error rate per syndrome extraction round.

\section{The biased erasure noise model} \label{app:BE}

A biased erasure is defined as a heralded exit from one half of the  computational subspace~\cite{grassl_codes_1997, sahay_high-threshold_2023}. The effective Kraus operators for this channel can be written as:
\begin{align}
\label{eq:w0kraus}
W_0 &= \ket{0}\bra{0} + \sqrt{1-2p_e} \ket{1}\bra{1} \\
W_e &= \sqrt{2 p_e} \ket{1}\bra{1}
\end{align}

This error channel is motivated by the metastable $^{171}$Yb atom qubit, where the dominant decay mechanism is decay from the Rydberg level $\ket{r}$ to a set of external states during two-qubit gates. Since the qubit is only excited to $\ket{r}$ from the $\ket{1}$ state during these gates,  detection of population in the external states is akin to measurement in the $Z$ basis with a known outcome. The qubit can be reinitialized by replacement with a fresh atom in $\ket{1}$. 

There are dual advantages to engineering qubits to experience dominant biased erasures as opposed to Pauli noise. Firstly, biased erasure are heralded errors, which means that the logical failure rate scales as $p_L \propto p^d$. In contrast, the only information available to identify qubits that have experienced Pauli errors within the computational subspace is their syndrome, leading to a scaling of $p_L \propto p^{(d+1)/2}$. Secondly, one benefits from the \textit{biased} nature of the erasure. If the error occurred during a CZ gate (or at the control of a CNOT gate), one can prove that effective operator on the qubit after atom replacement and Pauli twirling is $\left(I \rho I + Z \rho Z \right)/2$. If it occurred to the qubit at the target of a CNOT gate, the effective error channel is  $\left(I \rho I + X \rho X \right)/2$. Biased erasures thus allow identification of the apparent error ($X$ or $Z$) at a known location, lending to easier error correction and high thresholds.

For the metastable Yb, the theoretical maximum erasure fraction is $R_e =98\%$~\cite{wu_erasure_2022}. $R_e=33\%$ has recently been achieved experimentally via detection of leakage to the ground state~\cite{ma_high-fidelity_2023}.

\end{document}